\documentstyle[a4wide,amsfonts,cite]{article}

\newcommand{\vb}[1]{{\mathbf{#1}}}
\newcommand{\lb}[1]{\label{#1}}

\newcommand{\bc}{\begin{center}}
\newcommand{\ec}{\end{center}}
\newcommand{\be}{\begin{equation}}
\newcommand{\ee}{\end{equation}}
\newcommand{\bea}{\begin{eqnarray}}
\newcommand{\eea}{\end{eqnarray}}
\newcommand{\ba}[1]{\begin{array}{#1}}
\newcommand{\ea}{\end{array}}

\newcommand{\bt}[1]{\begin{table}[ht]\centering\begin{tabular}{#1}}
\newcommand{\et}[1]{\end{tabular}\caption{\small#1}\end{table}}

\begin{document}

\addtolength{\baselineskip}{0.20\baselineskip}

\thispagestyle{empty}

\begin{flushright} 
{\sl November 2007}\\
{\tt hep-ph/0609239}\\
\end{flushright}

\begin{center}

\vspace{1.5cm}

{\large\bf{A Pseudo-Photon in Non-Trivial Background Fields}}\\[15mm]

{\bf P. Castelo Ferreira}\\[5mm]
{\small{\it CENTRA, Instituto Superior T\'ecnico, Av. Rovisco Pais, 1049-001 Lisboa, Portugal}}\\[5mm]
{\small\tt pedro.castelo.ferreira@ist.utl.pt}\\[15mm]

{\bf Abstract}\\[3mm]
\begin{minipage}{13cm}
We show that in the presence of external fields for which either $\dot{\vb{B}}^{\mathrm{ext}}\neq 0$ or
$\nabla\times\vb{E}^{\mathrm{ext}}\neq 0$ it is not possible to
derive the classical Maxwell equations from an action with only one gauge field.
We suggest that one possible solution is to consider a second physical pseudo-vector
gauge field $C$. The action for this theory is originally motivated by the inclusion
of magnetic monopoles. These particles play no role in this
work and our argument is only based in, that the violation of the Bianchi identities, cannot be accounted
at the action level with only the standard gauge field. We give a particular example for a periodic rotating
external magnetic field. Our construction holds that at classical level both the vector and pseudo-vector
gauge fields $A$ and $C$ are regular. We compare pseudo-photon with paraphoton (graviphoton) theories
concluding that, besides the mechanisms of gauge symmetry breaking already studied,
the Bianchi identities violation are a crucial difference between both theories.
We also show that, due to Dirac's quantization condition, at quantum field theory level
the effects due to pseudo-photons and photons can be distinguished by the respective contributions
to the magnetic moment of fermions and vacuum polarization. These effects may be
relevant in astrophysical environments, namely close and inside neutron stars and magnetars.\\[5mm] \begin{center}\bf
Erratum\\[-3.5mm]\end{center} The construction in this work must, at most, be considered as a conceptual system or toy model. Are discussed systems where the
results may have relevance.\ \\\vfill
\end{minipage}
\end{center}
\vfill

\newpage
\section{Introduction and Conclusion}
The existence of a second gauge field (photon), as far as the author is
aware, have been first put forward in the context of electrodynamics
in the presence of magnetic monopoles~\cite{Dirac}, by Cabibbo and
Ferrari~\cite{CF} and further developed in proper-time formalism by
Schwinger~\cite{Schwinger_1}. Later Okun have proposed the existence of
massive vector fields called paraphotons~\cite{Okun}
(also known as graviphotons) which also imply the existence of millicharged fermions~\cite{Holdom}.
The proposal for a second physical pseudo-vector gauge field have further been developed in the context
of electromagnetic duality and theories including magnetic
monopoles~\cite{Sing_01,EB}.
The second proposal have been mostly applied to astrophysics~\cite{Okun,Holdom}
and more recently to vacuum polarization effects~\cite{paraphotons}.
The main difference between both theories are the transformation properties under the discrete
symmetries $P$ and $T$ of the physical degrees of freedom~\cite{Sing_01,EB}. Although
this seems simply a technicality it has very striking consequences in relation to the Bianchi identities,
mechanisms of gauge symmetry breaking and planar system physics~\cite{Proca_planar}.
The most easy way to explain the difference between both theories is to note that the map between them
presented in~\cite{EB} is generally non-regular, i.e. a pseudo-vector regular field configuration corresponds to a
vector field configuration that violates the Bianchi identities, hence non-regular.
In~\cite{EB} it has been shown that, upon the inclusion of magnetic charges and currents,
the singularities of the gauge field, either the Dirac string~\cite{Dirac} or
the Wu-Yang fiber bundle~\cite{WY}, can be removed by considering an extend $U_e(1)\times U_g(1)$
gauge symmetry. Here we go one step further, we show that even in the absence of magnetic and electric
charges such extended gauge symmetry is necessary to derive the standard classical Maxwell
equations in the presence of non-trivial background field configurations. We start our
demonstration by showing that, from the Maxwell action (with only the standard gauge
field $A$), it is not possible to derive the usual equations in the presence of non-trivial external fields obeying the
properties $\dot{\vb{B}}^{\mathrm{ext}}\neq 0$ or $\nabla\times\vb{E}^{\mathrm{ext}}\neq 0$.
External fields with either of these properties violate the Bianchi identities for the usual
gauge field $A$. Then we show that using the extended actions with two gauge fields
$A$ and $C$ introduced in~\cite{Sing_01,EB}, the correct Maxwell equations in the presence
of non-trivial background fields can be correctly derived from a variational principle.
We explicitly compute an example that coincides with the PVLAS experiment conditions~\cite{PVLA}
and discuss the implications for quantum field theory of our results, namely concerning
fermion magnetic moment and vacuum polarization due to Euler-Heisenberg~\cite{mark}.
For current accuracy of earth based optical experiments the effects presented here
are not measureable, however in astrophysical
environments, inside stars and near neutron stars and magnetars~\cite{magnetars},
where rotating strong magnetic fields are present, these effects are expected
to be relevant, for example for neutrino oscilations~\cite{neutrinos} and polarization
effects in gamma ray bursts~\cite{polarization}. Further we note that the
so far non-detection of pseudo photons is not necessarily a bad feature
of the theory, we recall that the standard gauge field $A$ was introduced much before experimental
evidence of its existence, that was only achieved in the early sixties due to the works
of Aharanov and Bohm~\cite{AB}.

\section{Maxwell Equations}
The standard Maxwell action in vacuum and in the absence of external sources is
\be
S_{\mathrm{Maxwell}}=-\frac{1}{4}\int dx^4 F_{\mu\nu}F^{\mu\nu}\ ,\ F_{\mu\nu}=\partial_\mu A_\nu-\partial_\nu A_\mu\ .
\lb{S_max}
\ee
The equations of motion with respect to the gauge field in covariant form are $\partial_\mu F^{\mu\nu}=0$ and
are supplemented by the Bianchi identities that hold only for regular gauge fields $\epsilon^{\mu\nu\delta\rho}\partial_\nu F_{\delta\rho}=0$.
Together these two equations correspond to the usual Maxwell equations. In vectorial form
we have, respectively the equations of motion and Bianchi Identities
\be
\partial_\mu F^{\mu\nu}=0\ \Leftrightarrow \ \left\{\ba{rcl}\displaystyle\nabla. \vb{E}&=&0\\[5mm]
                                                            \displaystyle\nabla\times \vb{B}&=&\displaystyle+\dot{\vb{E}}\ea\right.
\ ,\ 
\epsilon^{\mu\nu\delta\rho}\partial_\nu F_{\delta\rho}=0\ \Leftrightarrow \ \left\{\ba{rcl}\displaystyle\nabla. \vb{B}&=&0\\[5mm]
                                                                                   \displaystyle\nabla\times \vb{E}&=&\displaystyle-\dot{\vb{B}}\ \ea\right.
\lb{Maxwell}
\ee
In the presence of external applied fields the usual procedure is to decompose the fields into external (or applied) and internal (or induced)
components. The meaning of the components of such a decomposition is the following:
\begin{itemize}
\item \textit{External (or Applied) Fields} -- Stand for fields that are applied to the physical system by some
external mechanism. Here we use the notation
$E^{i,\mathrm{ext}}=\bar{F}^{0i}$ and $B^{i,\mathrm{ext}}=\epsilon^{ijk}\bar{F}_{jk}/2$.
\item \textit{Internal (or Induced) Fields} -- Stand for fields existing in the physical system which are not directly controllable by any external
mechanism. Here we use the notation $E^{i,\mathrm{ind}}$ and $B^{i,\mathrm{ind}}$ and are considering both internal photons ($a$ field),
internal pseudo-photons ($C$ field) and internal paraphotons ($\tilde{C}$ field).
\end{itemize}
Using the above definitions and notations we consider the gauge connection decomposition into external and internal fields
\be
F_{\mu\nu}=\bar{F}_{\mu\nu}+f_{\mu\nu}\ ,\ f_{\mu\nu}=\partial_\mu a_\nu-\partial_\nu a_\mu
\lb{decomp}
\ee
Taking directly the maxwell equations~(\ref{Maxwell}) we can solve them in the presence
of external fields and obtain the induced fields in vacuum. In this way the second set
of the Maxwell equations corresponding to the Bianchi identities will hold that, for an external applied magnetic field varying over time an
electric field will be induced, such that $\nabla\times \vb{E}^{\mathrm{ind}}=-\dot{\vb{B}}^{\mathrm{ext}}$. Also
for an external applied electrical field varying over space, a magnetic field will be induced such that
$\dot{\vb{B}}^{\mathrm{ind}}=-\nabla\times \vb{E}^{\mathrm{ext}}$. This is a standard result and there is plenty of
direct experimental evidence for it. It also clearly implies that we are violating the
Bianchi identities for the gauge field $a$ such that
\be
\epsilon^{\mu\nu\delta\rho}\partial_\nu f_{\delta\rho}=-\epsilon^{\mu\nu\delta\rho}\partial_\nu \bar{F}_{\delta\rho}\ .
\ee
Trying to deduce the same result from the action~(\ref{S_max}) by using the decomposition~(\ref{decomp}) is hopeless.
It is the Bianchi identities we are dealing with, we readily conclude that it is impossible. So if we want to
derive the same results from an action, both giving a more fundamental framework to this result and, simultaneously, allowing
for a quantum field theory treatment of it, we cannot get away with keep using the
Maxwell action~(\ref{S_max}). Other possible approaches to implement similar constructions
from a variational approach consist in considering extended Lagrangians where the gauge connection and the
gauge fields are considered independent variables~\cite{Deser}. These approaches also double
the degrees of freedom and the actions are explicitly invariant under electromagnetic duality which
is not, necessarily, a good feature~\cite{EB}. Here we will use an extended gauge symmetry $U_e(1)\times U_g(1)$
corresponding to two distinct internal gauge fields $a$ and $C$
\be
\ba{rcl}
S&=\displaystyle-\frac{1}{4}\int dx^4&\left[F_{\mu\nu}F^{\mu\nu}-G_{\mu\nu}G^{\mu\nu}-\epsilon^{\mu\nu\delta\rho}F_{\mu\nu}G_{\delta\rho}\right]\\[5mm]
 &=\displaystyle-\frac{1}{4}\int dx^4&\left[f_{\mu\nu}f^{\mu\nu}+\bar{F}_{\mu\nu}\bar{F}^{\mu\nu}+2\bar{F}_{\mu\nu}f^{\mu\nu}\right.\\[5mm]
 &                                   &\left.-G_{\mu\nu}G^{\mu\nu}-\epsilon^{\mu\nu\delta\rho}f_{\mu\nu}G_{\delta\rho}-\epsilon^{\mu\nu\delta\rho}\bar{F}_{\mu\nu}G_{\delta\rho}\right]\ ,
\ea
\lb{S}
\ee
where the new gauge connection in relation to the previous formulae is $G_{\mu\nu}=\partial_\mu C_\nu-\partial_\nu C_\mu$. This action was obtained in~\cite{EB} in order to give a variational description of electromagnetism that
consistently incorporates magnetic monopoles~\cite{Dirac} and, simultaneously, keep the gauge fields regular and is based in the
original works of Cabibbo and Ferrari~\cite{CF}. It is interesting that the external fields play the role of magnetic currents.
However, at the level we are working, monopoles do not play any role. For generic gauge fields (i.e. the gauge fields can be non-regular)
the equations of motion for $a$ and $C$ and the electric and magnetic field definitions are, respectively~\cite{Sing_01,EB},
\be
\left\{\ba{rcl}
\displaystyle\partial_\mu\left(f^{\mu\nu}-\frac{1}{2}\epsilon^{\mu\nu\delta\rho}G_{\delta\rho}\right)+\partial_\mu\bar{F}^{\mu\nu}&=&0\\[5mm]
\displaystyle\partial_\mu\left(G^{\mu\nu}+\frac{1}{2}\epsilon^{\mu\nu\delta\rho}f_{\delta\rho}\right)+\frac{1}{2}\epsilon^{\mu\nu\delta\rho}\bar{F}_{\delta\rho}&=&0
\ea\right.\ ,\ 
\left\{\ba{rcl}
E^i&=&\displaystyle f^{0i}-\frac{1}{2}\epsilon^{0ijk}G_{jk}\\[5mm]
B^i&=&\displaystyle G^{0i}+\frac{1}{2}\epsilon^{0ijk}f_{jk}
\ea\right.
\lb{EB_aC}
\ee
we obtain, in vectorial form, the standard Maxwell equations.
We further note that for regular gauge fields $a$ and $C$ the Bianchi identities
for each of them hold and the equations of motion decouple due to
$\epsilon^{\mu\nu\delta\rho}\partial_\nu G_{\delta\rho}=\epsilon^{\mu\nu\delta\rho}\partial_\nu f_{\delta\rho}=0$.
In the example that follows, the usual Maxwell equations hold with the definitions~(\ref{EB_aC}) and
the coupling between both gauge sectors is achieved through the external fields.

\section{An Example}
Here we are going to exemplify the above arguments and results by considering
a time-dependent magnetic field such that $\dot{B}^{\mathrm{ext}}\neq 0$ and $\vb{\nabla}\times \vb{E}^{\mathrm{ext}}=0$.
We will consider cylindrical symmetry around the $z$ axis such that we have a magnetic field
rotating with angular frequency $\omega_0$
\be
\ba{rcl}
\displaystyle\vb{B}^{\mathrm{ext}}(t)&=&\displaystyle B_0\left[\sin(\omega_0\,t),\cos(\omega_0\,t),0\right]\ ,\\[5mm]
\displaystyle\dot{\vb{B}}^{\mathrm{ext}}(t)&=&\displaystyle \omega_0\,B_0\left[\cos(\omega_0\,t),-\sin(\omega_0\,t),0\right]\ .
\ea
\lb{B_ext}
\ee
We can solve Maxwell equations~(\ref{Maxwell}) by recursion
\be
\left\{\ba{rcl}\nabla\times \vb{E}^{(1)}&=&-\dot{\vb{B}}^{(0)}\\[3mm] \nabla\times \vb{B}^{(1)}&=&+\dot{\vb{E}}^{(1)}\ea \right.
\ \ \ \ ,\ \ \ \ \ 
\left\{\ba{rcl}\nabla\times \vb{E}^{(n)}&=&-\dot{\vb{B}}^{(n-1)}\\[3mm] \nabla\times \vb{B}^{(n)}&=&+\dot{\vb{E}}^{(n)}\ea \right.\ \ ,
\ee
where $\vb{B}^{(0)}=\vb{B}^{\mathrm{ext}}$ stands for the external magnetic field~(\ref{B_ext}).
The full solutions of the electric and magnetic fields are
\be
\ba{rclcrcl}
\vb{E}&=&\vb{E}^{\mathrm{ind}}                      &\ ,\ \ \ \ &\vb{E}^{\mathrm{ind}}&=&\displaystyle\sum_{n=1}^\infty \vb{E}^{(n)}\ ,\\[5mm]
\vb{B}&=&\vb{B}^{\mathrm{ext}}+\vb{B}^{\mathrm{ind}}&\ ,\ \ \ \ &\vb{B}^{\mathrm{ind}}&=&\displaystyle\sum_{n=1}^\infty \vb{B}^{(n)}\ ,
\ea
\ee
such that the induced field solutions are
\be
\ba{rcl}
\vb{E}^{\mathrm{ind}}&=&\displaystyle B_0\left[0,0,-\sin(\omega_0\,y)\cos(\omega_0\,t)-\sin(\omega_0\,x)\sin(\omega_0\,t)\right]\\[5mm]
\vb{B}^{\mathrm{ind}}&=&\displaystyle B_0\left[(\cos(\omega_0\,y)-1)\sin(\omega_0\,t),(\cos(\omega_0\,x)-1)\cos(\omega_0\,t),0\right]\ .
\ea
\lb{sol_EB}
\ee
Given these solutions we can go back to the equations of motion~(\ref{EB_aC}) and using
the field definitions~(\ref{EB_aC}) write the Bianchi identities for both
the exterior vector field $A$ (external photon), the internal vector field $a$ (internal photon),
the internal pseudo-vector field $C$ (pseudo-photon) and the equivalent vector field $\tilde{C}$ (paraphoton):
\be
\ba{lrcl}
\mathrm{external\ photon:}&\nabla\times \vb{E}_A^{\mathrm{ext}}+\dot{\vb{B}}_A^{\mathrm{ext}}&\neq&0\ ;\\[5mm]
\mathrm{internal\ photon:}&\nabla\dot\vb{E}_a=\nabla\times \vb{E}_a+\dot{\vb{B}}_a&=&0\ ;\\[5mm]
\mathrm{pseudo-photon:}\ \ &\nabla\dot\vb{E}_C=\nabla\times \vb{E}_C+\dot{\vb{B}}_C&=&0\ ;\\[5mm]
\mathrm{paraphoton:}&\nabla\times \vb{E}_{\tilde{C}}+\dot{\vb{B}}_{\tilde{C}}&=&\dot{\vb{B}}^{\mathrm{ext}}\neq 0\ .
\ea
\ee
The definitions used for the electric and magnetic fields corresponding to the several gauge fields
$A$, $a$, $C$ and $\tilde{C}$ are the same of~\cite{EB} and we fully list them:
\be
\ba{lrclcrcl}
\mathrm{external\ photon:}&E_A^{i,\mathrm{ext}}&=&\bar{F}^{0i}&\ ,\ \ &B_A^{i,\mathrm{ext}}&=&\displaystyle\frac{1}{2}\epsilon^{ijk}\bar{F}_{jk}\ ;\\[5mm]
\mathrm{internal\ photon:}&E_a^i&=&f^{0i}&\ ,\ \ &B_a^i&=&\epsilon^{ijk}f_{jk}\ ;\\[5mm]
\mathrm{pseudo-photon:}&E_C^i&=&\displaystyle-\frac{1}{2}\epsilon^{ijk}G_{jk}&\ ,\ \ &B_C^i&=&G^{0i}\ ;\\[5mm]
\mathrm{paraphoton:}&E_{\tilde{C}}^i&=&\tilde{G}^{0i}&\ ,\ \ &B_{\tilde{C}}^i&=&\displaystyle\frac{1}{2}\epsilon^{ijk}\tilde{G}_{jk}\ .
\ea
\ee
Here we use the same definitions for the connections as given in~(\ref{S_max}),~(\ref{decomp}),~(\ref{S}) and the
map between the pseudo-photon $C$ and the paraphoton $\tilde{C}$ is generally given in terms of the gauge connections~\cite{EB}
\be
\tilde{G}^{\mu\nu}=\frac{1}{2}\epsilon^{\mu\nu\delta\rho}G_{\delta\rho}\ .
\lb{dual}
\ee
This equation maps regular $C$ into non-regular $\tilde{C}$. We have just presented such an example. 
Therefore we conclude that the violation of the Bianchi identities is one crucial difference between
pseudo-photon and paraphoton theories, only pseudo-photons allow to fully describe the classical
Maxwell equations maintaining the gauge fields regular. This fact is due to the Hopf term in
action~(\ref{S}) that, as already extensively explained in~\cite{EB},
explicitly couples the charge of one gauge group to the topological charge of the
other gauge group.

\section{Potential Detectability of Pseudo-Photons}
It remains to discuss for which physical systems the results presented in this work
may be relevant. In particular we address here the coupling of pseudo-photons to
Fermions at quantum field theory level. In~\cite{Proca_planar} it was introduced a covariant derivative such that for
external field configurations as the ones presented here for which $\tilde{C}\neq 0$
we expect the Lagrangian corresponding to the Dirac fermions coupling to the gauge fields to be
\be
{\mathcal{L}}_\psi=\bar{\psi}\left(i\gamma^\mu\partial_\mu-e\gamma^\mu A_\mu-g\gamma^\mu\tilde{C}_\mu-m\right)\psi\ .
\ee
Here we use complex notation, the $\gamma^\mu$ matrices are the usual Dirac matrices
and $\bar{\psi}=\psi^\dag\gamma^0$. If this coupling exist, due to Dirac quantization condition~\cite{Dirac}
we will obtain a quantitative difference between the usual coupling to background fields
due to the standard photon $A$ and the pseudo-photon $C$ given in terms
of the dimensionless fine-structure constants and dimensionless unit charges ratios by
\be
\alpha_e\alpha_g=\frac{e^2}{4\pi\epsilon_0\hbar c}\,\frac{g^2}{4\pi\mu_0\hbar c}=\frac{n^2}{4}\ \ ,\ \ \frac{\alpha_g}{\alpha_e}\approx 4692.2\ \ ,\ \ \sqrt{\frac{\epsilon_0}{\mu_0}}\,\frac{g}{e}\approx 68.5\ ,
\ee
where we used the vacuum resistance defined in terms of the vacuum
dielectric constant $\epsilon_0$ and magnetic permeability $\mu_0$
and Dirac's quantization condition with $n=1$~\cite{Dirac} in IS unit, which we adopt in the remaining of this work.
The consequences of this result is that the fermion magnetic moment for the coupling to
external fields due to the pseudo-photon ($\mu_C$) is larger by almost two orders of magnitude in relation
to the coupling due to the photon ($\mu_A$), for example for the electron we have that
\be
\mu_A=\frac{e}{m_e}\ \ , \ \ \mu_C=\sqrt{\frac{\epsilon_0}{\mu_0}}\frac{g}{m_e}\approx 68.5\times \mu_A\ .
\ee
By considering an expansion in the radius $r=\sqrt{x^2+y^2}$ for the induced magnetic field~(\ref{sol_EB}) and comparing
it with the external magnetic field~(\ref{B_ext}) we obtain that the magnetic moment effects due to pseudo-photons
become more relevant than the magnetic moment effects of photons for
\be
r>\frac{c}{68.5\,\omega_0}\sim\frac{4.37\times 10^6}{\omega_0}\ \ \ (meters)
\lb{bound_mu}
\ee
Also in terms of vacuum polarization in the presence of external fields we have significant changes,
for example the radiative corrections due to one-loop QED corrections is given by the Euler-Heisenberg Lagrangian~\cite{HE}
\be
{\mathcal{L}}^{(2)}_e=-\xi_e \left(4({\mathcal{F}}_{\mu\nu}{\mathcal{F}}^{\mu\nu})^2+7(\epsilon^{\mu\nu\delta\rho}{\mathcal{F}}_{\mu\nu}{\mathcal{F}}_{\delta\rho})^2\right)\ ,\ \ \ \ \xi_e=\frac{2\hbar^3\epsilon_0}{45\,m_e^4c^5}\alpha_e^2\ .
\lb{actione}
\ee
Here we are using the generalized definition for the gauge connection
as introduced in~\cite{Sing_01,EB}, i.e. ${\mathcal{F}}^{\mu\nu}=F^{\mu\nu}-\sqrt{\frac{\epsilon_0}{\mu_0}}\,\frac{g}{e}\epsilon^{\mu\nu\delta\rho}G_{\delta\rho}/2$
and the respective definition of the electric and magnetic fields given by~(\ref{EB_aC}).
The non-linear contribution to the radiation equation for the example discussed in the previous sections hold the dispersion
relations~\cite{HE}
\be
\omega_\pm=k\left(1-\lambda^e_\pm Q_0^2-\lambda^g_\pm Q_{\mathrm{ind}}^2)\right)\ \ ,\ \ \lambda^g_\pm=\frac{\alpha_g^2}{\alpha_e^2}\lambda^e_\pm \approx 2.2\times 10^7\lambda^e_\pm\ .
\ee
Here we have $\lambda^e_+=7\xi_e$ and $\lambda^e_-=4\xi_e$ and $Q_0^2$ and $Q_{\mathrm{ind}}^2$ are respectively the
contribution from the usual background fields and the induced background fields due to the pseudo-photon. For
the example developed in the previous sections we obtain
\be
Q_0^2=B_0^2\ ,\ \ \ Q_{\mathrm{ind}}^2= B_0^2\left[\left(\cos(\omega_0\,x)-1\right)^2+\left(\cos(\omega_0\,y)-1\right)^2\right]\ .
\ee
From~(\ref{sol_EB}) and~(\ref{B_ext}) we obtain that the effect due to pseudo-photons becomes more relevant than
the effect due to photons for
\be
r>\frac{c}{\omega_0}\sqrt{\frac{1}{2.2\times 10^7}}\sim \frac{6.4\times 10^4}{\omega_0}\ \ \ (meters)
\lb{bound_HE}
\ee
None of these effects have so far been observed experimentally and are out of the current accuracy of
earth based optical experiments~\cite{harm}, in PVLAS experiment~\cite{PVLA} the standard vacuum polarization is of order
$I/I_0\sim 10^{-22}$ and from the above result, considering $r\sim 10^{-3}\ m$ we obtain $I/I_0\sim 10^{-39}$.
However inside and near neutron stars and magnetars~\cite{magnetars} the effects presented in this work may be significant,
for these astrophysical environments magnetic fields go up to $10^{12}\ T$ with rotations
of up to $10^3\ Hz$ which satisfy the bounds~(\ref{bound_mu}) and~(\ref{bound_HE})
for radius superior to $4.4\ Km$ and $63.9\ m$ affecting, for example, neutrino magnetic moment~\cite{neutrinos}
and gamma-ray bursts polarization~\cite{polarization}. The topics covered in this section will
be fully developed somewhere else.\\[-2mm]

\noindent {\large\bf Acknowledgements}\\
The author thanks J. T. Mendon\c{c}a for very useful discussions and motivation in the course of this work
as well as to the referee for valuable suggestions that significantly improved the manuscript.
This work was supported by SFRH/BPD/17683/2004.\\[-4mm]

\appendix

\section{Erratum}

In~\cite{article} it was considered a rotating magnetic field $B^i=\epsilon^{ijk}\partial_j A_k$ that violates
the Bianchi identities for the standard photon (i.e. the vector gauge field $A$). It is claimed by the author that this
construction justifies the existence of a physical pseudo-photon (a pseudo-vector gauge field $C$)
independently of the detectability of magnetic monopoles~\cite{monopoles}. This is an overstatement, in particular
the author failed to notice that the construction presented in~\cite{article} is, in standard electromagnetism,
unphysical. Considering the usual Maxwell equations in vacuum a magnetic field does not exist by itself, it needs
to be generated by some source. Usually an electric current or its equivalent (e.g a permanent magnet)~\cite{Jackson}.
By properly considering the electric currents that generate the magnetic field $B^i$ in the Maxwell action the standard
vector field $A$ is enough to describe most physical systems and generally has regular solutions.

Hence, although the calculations are correct and the derivations presented in~\cite{article} are consistent with the
assumption that a single rotating magnetic field may exist, the simple example carried out must, at most, be considered
as a conceptual system or a toy model. Nevertheless it is important to stress that this simple example may have relevance
in some particular practical implementations when considering magnetic flux tubes or strings.

As examples there are the Abrikosov-Nielsen-Olesen string solutions~\cite{strings}. Besides
being able to reproduce the confinement of magnetic monopoles~\cite{Nambu} are relevant in systems such
as type~II superconductivity. Although laboratory electron systems are usually not rotating, a more suitable framework
where the same solutions may exist are neutron stars and pulsars (these system have already been mentioned in~\cite{article})
for which a neutron type~II superconductivity phase could be present~\cite{neutronE}.

Also for standard stars as the Sun has been put forward within magneto-hydrodynamics~\cite{Parker} that singular magnetic field
lines (magnetic flux tubes thinner than the plasma Debye length) may explain the heating of the stellar
corona due to magnetic reconnection mechanisms~\cite{Priest}. In this framework are considered topological defects and topology
changing mechanisms in the stellar plasma. In particular it is interesting to note that, in planar magnetic reconnection theory,
the magnetic fields lye along the plane of the plasma while the electric fields are orthogonal to this plane~\cite{Priest}.
It is important to stress that such fields are not allowed in the standard planar Maxwell theory where only planar electric fields
and orthogonal magnetic fields exist. These results are consistent with the field content obtained in planar systems when
pseudo-photon fields are considered~\cite{planar}, hence seems relevant to consider the existence of pseudo-photons in these frameworks.\\[-2mm]

\noindent {\large\bf Acknowledgments}\\
The author thanks Fernando Barbero for several discussions, in particular by
noting that in standard electromagnetic systems a non-regular magnetic field
is unphysical as well as thanks the hospitality of the CSIC in Madrid. Also thanks Mar
Bastero-Gil for discussions and the hospitality of HEP group of the University
of Granada as well as to Jo\~ao Pulido, Jorge Dias de Deus, Pedro Sacramento,
D\'ario Passos and Il\'{\i}dio Lopes for discussions. This work was supported by
SFRH/BPD/17683/2004 and SFRH/BPD/34566/2007.\\[-4mm]


\begin{thebibliography}{99}

\bibitem{Dirac} P. A. M. Dirac, Proc. Roy. Soc. {\bf A133} (1931) 60; Phys. Rev. {\bf 74} (1948) 817.

\bibitem{CF} N. Cabibbo and E. Ferrari, Il Nuovo Cimento {\bf XXIII} No 6 (1962) 1147.

\bibitem{Schwinger_1} J. Schwinger, Phys. Rev. {\bf 173} (1968) 1536.

\bibitem{Okun} L. B. Okun, Sov. Phys. JETP {\bf 56} (1982) 502; Zh. Eksp. Teor. Fiz. {\bf 83} (1982) 892-898.

\bibitem{Holdom} B. Holdom, Phys. Lett. {\bf B166} (1986) 196; Phys. Lett. {\bf B178} 65-70; S. Davidson and M. Peskin, Phys. Rev. {\bf D49} (1994) 2114.

\bibitem{Sing_01} D. Singleton, Int. J. Theor. Phys. {\bf 34} (1995) 2453, \texttt{hep-th/9701040}; Am. J. Phys. {\bf 64} (1996) 452; P. C. R. Cardoso de Mello, S. Carneiro and M. C. Nemes, Phys. Lett. {\bf B384} 197, \texttt{hep-th/9609218}; N. Berkovits, Phys. Lett.  {\bf B395} (1997) 28-35, \texttt{hep-th/9610134}.

\bibitem{EB} P. Castelo Ferreira, J. Math. Phys. {\bf 47} (2006) 072902, \texttt{hep-th/0510063}.

\bibitem{paraphotons} M. Gasperini, Phys. Lett. {\bf B263} (1991) 267-269; Phys. Lett. {\bf B237} (1991) 431; R. Cameron et al., Phys. Rev. {\bf D47} (1993) 3707-3725; E. Mass\'o and J. Redondo, JCAP {\bf 0509} (2005) 015,  \texttt{hep-ph/0504202}; Phys. Rev. Lett. {\bf 97} (2006) 151802, \texttt{hep-ph/0606163}; E. Mass\'o, \texttt{hep-ph/0607215}; B. Batell, T. Gherghetta, Phys. Rev. {\bf D73} (2006) 045016, \texttt{hep-ph/0512356}; H. Gies, J. Jaeckel and A. Ringwald, Phys. Rev. Lett. {\bf 97} (2006) 140402, \texttt{hep-ph/0607118}; Europhys. Lett. {\bf 76} (2006) 794-800, \texttt{hep-ph/0608238}; S. A. Abel, J. Jaeckel, V. V. Khoze and A. Ringwald, \texttt{hep-ph/0608248}.

\bibitem{Proca_planar} P. Castelo Ferreira and J. T. Mendon\c{c}a, \texttt{hep-th/0601171}; J. T. Mendon\c{c}a and P. Castelo Ferreira, Europhys. Lett. {\bf 75} 189, \texttt{hep-th/0601166}; P. Castelo Ferreira, europhys. Lett {\bf 79} (2007) 20004, \texttt{hep-th/0703193}; \texttt{hep-th/0703194}.

\bibitem{WY} T. T. Wu and C. N. Yang, Phys. Rev. {\bf D12} (1975) 3845; Phys. Rev. {\bf D14} (1976) 437.

\bibitem{PVLA} E. Zavattini and al., Phys. Rev. Lett. {\bf 96} (2006) 110406, \texttt{hep-ex/0507107}.

\bibitem{mark} M. Marklund, P. K. Shukla, Rev. Mod. Phys. {\bf 78} (2006) 591, \texttt{hep-ph/0606099}.

\bibitem{magnetars} R. C. Duncan and C. Thompson,  Astrophys. J. {\bf 392} (1992) 9-13; C. Kouveliotou et al., Nature {\bf 393} (1998) 235-237; J. M. Lattimer and M. Prakash, Science {\bf 304} (2004) 536-542; P. Kaaret and al., \texttt{astro-ph/0611716}.

\bibitem{neutrinos} A. Cisneros, Astrophys. Space Sci.10 (1971) 87-92; L.B. Okun, M.B. Voloshin, M.I. Vysotsky, Sov. Phys. JETP {\bf 64} (1986) 446-452; Zh. Eksp. Teor. Fiz. {\bf 91} (1986) 754-765; M. B. Voloshin, Sov. J. Nucl. Phys. {\bf 48} (1988) 512; Yad. Fiz. {\bf 48} (1988) 804-810; S. M. Barr, E.M. Freire and A. Zee, Phys. Rev. Lett. {\bf 65} (1990) 2626; J. Pulido, Phys. Rept. {\bf 211} (1992) 167-199.

\bibitem{polarization} D. Bersier and al, Astrophys. J. {\bf 583} (2003) L63-L66, \texttt{astro-ph/0206465}; W. Coburn and S. E. Boggs, Nature {\bf 423} (2003) 415, \texttt{astro-ph/0305377}; J. Granot, Astrophys. J. {\bf 596} (2003) L17-L21.

\bibitem{AB} Y. Aharonov and D. Bohm, Phys. Rev. {\bf 115} No 3 (1959) 485.

\bibitem{Deser} S. Deser and O. Sarioglu, Phys. Lett. {\bf B423} (1998) 369-372, \texttt{hep-th/9712067}.

\bibitem{HE} W. Heisenberg and H. Euler, Z. Physik {\bf 98} (1936) 714, \texttt{physics/0605038}; J. Schwinger, Phys. Rev. {\bf 82} (1951) 664; Z. Bialynicka-Birula and I. Bialynicka-Birula, Phys. Rev. {\bf D2} (1970) 2341; S. L. Adler, Ann. Phys. {\bf 67} (1971) 599.

\bibitem{harm} J. T. Mendon\c{c}a, J. Dias de Deus and P. Castelo Ferreira, Phys. Rev. Lett. {\bf 97} (2006) 100403; {\bf 97} (2006) 269901(E), \texttt{hep-ph/0606099}; S. L. Adler, J. Phys. {\bf A40} (2007) F143-F152, hep-ph/0611267; S. Biswas, K. Melnikov, Phys. Rev. {\bf D75} (2007) 053003, \texttt{hep-ph/0611345}.

\end{thebibliography}

\begin{thebibliography}{99}

\bibitem[A1]{article} P. Castelo Ferreira, Phys. Lett. {\bf B651} (2007) 74-78, \texttt{hep-ph/0609239}

\bibitem[A2]{monopoles} P. A. M. Dirac, Proc. Roy. Soc. {\bf A133} (1931) 60; Phys. Rev. {\bf 74} (1948) 817; N. Cabibbo and E. Ferrari, Il Nuovo Cimento {\bf XXIII} No 6 (1962) 1147; P. Castelo Ferreira, J. Math. Phys. {\bf 47} (2006) 072902, \texttt{hep-th/0510063}.

\bibitem[A3]{Jackson} J. D. Jackson, \textit{Classical Electrodynamics}, $2^{nd}$ Edition, John Wiley \& Sons, 1975.

\bibitem[A4]{strings} A. A. Abrikosov, Sov. Phys. JETP {\bf 5} (1957) 1174-1182; Zh. Eksp. Teor. Fiz. {\bf 32} (1957) 1442-1452; H. B. Nielsen and P. Olesen, Nucl. Phys. {\bf B61} (1973) 45.
\bibitem[A5]{Nambu} Y. Nambu, \textit{String, Monopoles, and Gauge Fields}, Phys. Rev. {\bf D10} (1974) 4262.

\bibitem[A6]{neutronE} E. Flowers, M. Ruderman and P. Sutherland, Astrophys. J. {\bf 205} (1976) 541; D. Bailin and A. Love, J. Phys. {\bf A12} (1979) L283; Phys. Rept. {\bf 107} (1984) 325; J. A. Harvey, M. A. Ruderman and J. Shaham, Phys. Rev. {\bf D33} (1986) 2084; D. G. Yakovlev, K. P. Levenfish and Y. A. Shibanov, Phys. Usp. {\bf 169} (1999) 825; Astron. Lett. {\bf 25} (1999) 417; U. Lombardo and H. J. Schulze, Lect. Notes Phys. {\bf 578} (2001) 30-53; D. J. Dean and M. Hjorth-Jensen, Rev. Mod. Phys.{\bf 75} (2003) 607-656; I. Wasserman, Mon. Not. Roy. Astron. Soc. {\bf 341} (2003) 1020; D. Page, J. M. Lattimer, M. Prakash and A. W. Steiner, Astroph. J. Supp. {\bf 155} (2004) 623; D. Page, U. Geppert and F Weber, Nucl. Phys. {\bf A777} (2006) 497-530; S. Popov, H. Grigorian and D. Blaschke, Phys. Rev. {\bf C74} (2006) 025803.

\bibitem[A7]{Parker} E. N. Parker, \textit{The Solar Magnetic Field}, Proceedings of the VI Canary Islands School, Ed. T. Roca Cortes and F. Sanchez, Cambridge University Press 1996 p. 300.

\bibitem[A8]{Priest} E. R. Priest, \textit{Solar Magnetohydrodynamics}, Kluwer Academic Publishers 2000; Astroph. and Sp. Sci. {\bf 264} (1999) 77; E. Priest and T. G. Forbes, \textit{Magnetic Reconnection}, Cambridge University Press 2000; D. Biskamp, \textit{Magnetic Reconnection in Plasmas}, Cambridge University Press 2000.

\bibitem[A9]{planar} P. Castelo Ferreira, Europhys. Lett. {\bf 79} (2007) 20004, \texttt{hep-th/0703193}.



\end{thebibliography}
\end{document}